\documentclass[aps,prf,onecolumn,superscriptaddress,amsmath,amssymb]{revtex4-2}
\usepackage{amsmath}
\usepackage{mathtools}
\usepackage{amssymb}
\usepackage{color}
\usepackage{graphics}
\usepackage{epsfig}
\usepackage[normalem]{ulem} %to strike the words
\usepackage{cancel}
\usepackage[mathlines]{lineno}
\graphicspath{{Figures/}}
\usepackage{xtab,afterpage,longtable}
\usepackage{booktabs}   
\usepackage{ltablex}
\usepackage{float}
\relax

\begin{document}
\title{Intermittent Viscoelastic Turbulence in Strongly Coupled Plasmas
}
\author{Rauoof Wani}
\email{abrauoofwani@gmail.com}
\affiliation{Indian Institute of Technology Jammu, Department of Physics, Jammu, 181221, India}
% \author{Mahendra Verma}
% \affiliation{Department of Physics, Indian Institute of Technology Kanpur, Kanpur, 208016, India}
\author{Sanat Tiwari}
\affiliation{Indian Institute of Technology Jammu, Department of Physics, Jammu, 181221, India}
\date{\today}
%%%%%%%%%%%%%%%%%%%%%%%%%%%%%%%%%%%%%%%%%%%%%%%%%%%%%%
\begin{abstract}
Turbulence in strongly coupled plasmas (SCP) poses a unique challenge due to long-range interparticle interactions that impart viscoelastic properties to the medium. We report the first observation of intermittent viscoelastic turbulence in such plasmas using large-scale three-dimensional molecular dynamics simulations. In driven dissipative SCP, we observe scale-dependent flow dynamics arising from the interplay between potential (elastic) and viscous dissipation at the particle level.
Unlike conventional turbulence, governed by inertia and viscosity, SCP exhibits distinct energy transfer processes due to its intrinsic viscoelasticity. Both kinetic and elastic energy spectra show power-law scaling $E(k), \Phi(k) \propto k^{-3.5}$. In real space, velocity structure functions display nontrivial scaling, and the probability distribution functions of velocity increments deviate significantly from Gaussian behavior, especially in the tails, signatures of intermittency. These findings establish SCP as a novel platform for studying viscoelastic turbulence, with broad relevance to astrophysical plasmas, soft matter, and nonequilibrium statistical physics.
\end{abstract}
%%%%%%%%%%%%%%%%%%%%%%%%%%%%%%%%%%%%%%%%%%%%
\maketitle
\label{intro_RTI}
%%%%%%%%%%%%%%%%%%%%%%%%%%%%%%%%%%%%%%%%%%%%
\section{introduction}
Hydrodynamic turbulence typically occurs in flows with high Reynolds number, $Re\sim O(10^3)$, where inertial stresses dominate over viscous stresses~\cite{Kolmogorov:DANS1941Structure,Kolmogorov:DANS1941Dissipation,Frisch:book}. However, turbulence can also emerge at much lower $Re\sim O(1)$ when the Weissenberg number $Wi$, representing the ratio of elastic to viscous stresses, exceeds unity~\cite{Groisman:Nature2000,groisman2001efficient}. This phenomenon, known as elastic turbulence, is characteristic of viscoelastic fluids such as polymeric solutions, gels, and biological fluids like blood~\cite{Dallas:PRE_2010,datta2022perspectives,wensink2012meso}. Under external stress, these systems store a significant portion of  energy elastically; once the stress is removed, the stored energy induces anisotropy and modifies energy transfer across scales~\cite{singh2024intermittency}.

Elastic turbulence has gained considerable attention over the past few decades, with numerical and theoretical studies often based on Oldroyd-B or FENE-P models highlighting the role of elastic instabilities and polymer stretching in driving chaotic motion~\cite{Berti:PRE2010,berti2008two}. Recent work has also examined its onset, transition mechanisms~\cite{varshney2019elastic}, and implications for microfluidic transport and mixing. Yet, the fundamental nature of viscoelastic turbulence remains elusive, lacking universal scaling laws. Here, we show that strongly coupled plasmas (e.g., dusty plasmas) provide an excellent test bed for probing viscoelastic turbulence at the microscopic level. Our results further reveal that turbulence in SCPs is marked by pronounced intermittency rather than smooth behavior.

Turbulence in SCPs is crucial for understanding complex flow dynamics in astrophysical settings, laboratory dusty plasma experiments, and high-energy density systems~\cite{fortov2005complex,Shukla_pk}. Experimental and numerical studies, particularly in two-dimensional dusty plasma liquids have revealed rich turbulent features such as coherent vortex structures, energy cascades, and wave driven turbulence, observed directly at the level of individual dust grains~\cite{lin2018interacting,Sachin_PoP_2024,Tsai_PRE_2014,sanat_ghz_turbulence,E_joshi_PRR,Rauoof_PoP_2024}. While significant progress has been made in understanding flow dynamics, viscoelastic turbulence, chaotic flows driven by elastic stresses, commonly seen in polymeric fluids has not yet been reported in these systems. Here, we show that dusty plasmas can indeed exhibit viscoelastic turbulence and intermittent flow behavior when the interparticle potential energy becomes comparable to thermal kinetic energy, even at low Reynolds numbers.

The viscoelastic effects in SCP arise naturally due to particle-level interactions~\cite{feng2010viscoelasticity,AshwinJoy_PRL_2005,Donko_PRL_2006}. The particles in the medium interact via pairwise interaction potential. Depending upon the coupling strength $\Gamma$ (which is the ratio of Coulomb potential energy to thermal energy between the particles), the medium shows phase transitions and intermediate states as the strength of interparticle interaction exceeds to the thermal energy~\cite{Murillo_2014_coupling,fortov2006physics}. Strongly coupled dusty plasmas (present case study) generally fall in a low Reynolds number regime (~1–100)~\cite{Sachin_PoP_2024} and have been observed to exhibit shear thinning, elastic wave propagation a characteristic feature of non-Newtonian fluids~\cite{saigo2002shear,Paramik_PRL}. This leads to a fundamental question: Can SCP exhibit viscoelastic turbulence, and if so, how is energy exchanged among the different modes of the system.? To investigate this, we employ three-dimensional (3D) molecular dynamics (MD) simulations to model the dusty plasma as a case study. MD offers a foundational and powerful framework for investigating both microscopic and macroscopic dynamics in SCP systems, as it inherently captures strong interparticle correlations and accurately resolves transport properties without the need for empirical approximations~\cite{Wani_2022,WANI2024129944,Rauoof_PoP_2024}. 

\section{Methodology and Simulation Details}
We simulate a dusty plasma as a 3D periodic system of two million particles interacting via pairwise Debye–Hückel potential. Simulations are performed using the open-source molecular dynamics package LAMMPS~\cite{lammps_thompson}, within a cubic domain $lx=ly=lz$ with periodic boundary conditions. The particle trajectories $\mathbf{r}_i(t)$ and velocities $\mathbf{v}_i(t)$ are calculated by integrating the classical equations of motion along with the thermal heat bath:
\begin{align}
\frac{d^2\mathbf{r}_i}{dt^2} &= - \frac{1}{m_i} \nabla \sum_{i,j} \phi_{ij} - \xi \mathbf{v}_i \\
\frac{d\xi}{dt} &= \frac{1}{gk_BT\tau^2} \left( \sum_i m_i \mathbf{v}_i^2 - g k_B T \right)
\end{align}

where $\phi(r)=(q^2/4\pi\epsilon_0r)e^{-r/\lambda_D}$, is the interaction potential between the particles, $a$, and $\lambda_D$ are the average separation between particles and the Debye length of background plasma respectively. $\xi$ is a friction coefficient, $\tau$ is a thermostat time constant, $g$ is the number of degrees of freedom, and $T$ is the temperature of the system. Initially, the multiparticle system is maintained at a fixed temperature and isolated from external influences to ensure total energy conservation. At equilibrium, the system achieves the target temperature corresponding to the coupling parameter $\Gamma$, with temperature and energy fluctuations maintained below $10^{-4}\%$ and $10^{-2}\%$, respectively. 
The details of the simulation and plasma parameters are provided in Table~\ref{sim_para}. 
%and \ref{pl\_para}.
%%%% Table-1
\begin{table}
\centering
\caption{Simulation parameters}
{\renewcommand{\arraystretch}{1.5}
\begin{tabular}{|l|l|} 
\hline
Number of particles, N &
$2 \times 10^6$
\\ 
\hline
Particle mass, m &
$6.9 \times 10^{-13}\mathrm{Kg}$
\\ 
\hline
Number density, n      & 
$6.69 \times 10^9 \mathrm{m^{-3}}$
\\ 
\hline
Particle charge, q     &  
$15 \times 10^3$e  
\\
\hline
Screening parameter, $\kappa$ $(a/\lambda_D)$    &  
0.1 
\\
\hline
Coupling strength, $\Gamma$ $(q^2/4\pi \epsilon_0 a k_BT$)   &  
1
\\
\hline
System length, $l_x=l_y$    &  
$ 200a $
\\
\hline
Dust frequency, $\omega_{pd}$ $(\sqrt{(ne^2/\epsilon_0 m)})$     &  
$ 79.2~\mathrm{Hz} $
\\
\hline
\end{tabular}
}
\label{sim_para}
\end{table}

The steady state system is achieved by continuously injecting energy in the form of 3D Taylor-Green vortex flow which provides a smooth and divergence-free initial condition that naturally evolves into turbulence through nonlinear interactions, it has a mathematical form
\begin{align}
\nonumber   
u(x, y, z, 0) &= U_0 \sin(x) \cos(y) \cos(z) \\
\nonumber
v(x, y, z, 0) &= -U_0 \cos(x) \sin(y) \cos(z) \\
\nonumber
w(x, y, z, 0) &= 0.
\end{align}
A thermal bath is attached to regulate the system temperature. The flow velocity $U_0=0.1v_{th}$ is kept an order of magnitude lower than the system thermal velocity $(v_{th})$ to ensure incompressibility of the system with Mach number $(M_a=0.1)$. It takes almost 500$\omega_{pd}^{-1}$ plasma periods for the system to achieve the steady state corresponding to coupling strength $\Gamma=1$, and this time period increases with an increase in $\Gamma$ because of the high correlation between the particles.
\begin{figure}[!ht]
    \centering
    \includegraphics[width=0.5\textwidth]{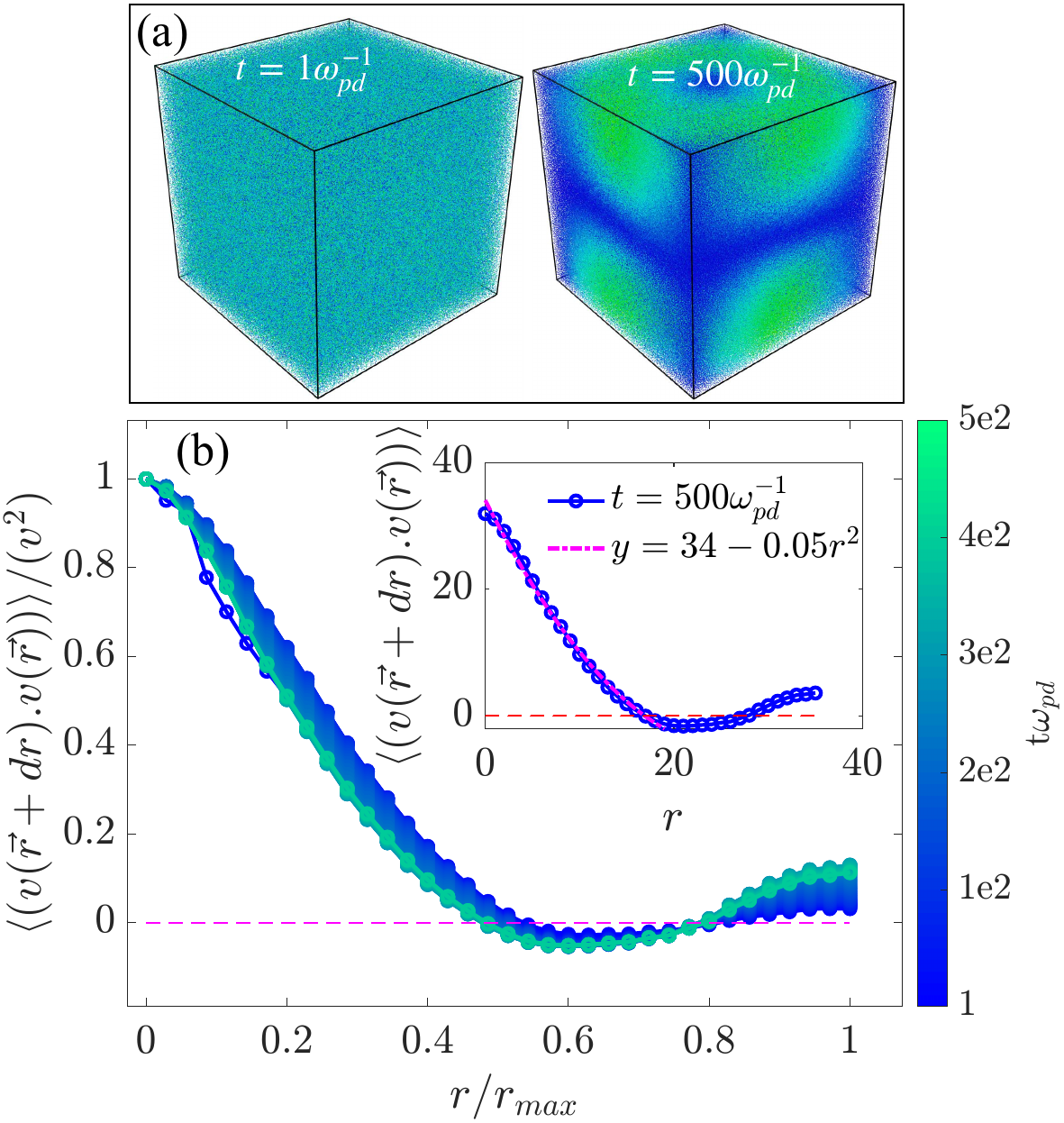}
 \caption{
   (a) Snapshots of particle trajectories at $t=1$ and $t=500$ normalized time units in a SCP system corresponding to $\Gamma=1$. The large scale Taylor-Green vortex can be seen dominated at 500$\omega_{pd}^{-1}$ where steady state is achieved by the system. (b) The evolution of velocity correlations and their slow decay showing some power law scaling (inset), which ensures the long lived coherent structure in the system.
    }
    \label{fig:TGV_evolution}
\end{figure}

During the initial stages, the damping dominates the external energy input causing the structure to diffuse. As time progresses, the continuous energy input strengthens the structure gradually overcoming the damping. At $t=500\omega_{pd}^{-1}$, (fig.~\ref{fig:TGV_evolution}(a)) as the system approaches steady state, large-scale structures begin to dominate, reflecting the characteristic scale of energy injection. The spatial velocity correlations decay slowly with power law scaling, indicating the long-lived coherent structures in the system. The spatial velocity correlations (fig.~\ref{fig:TGV_evolution}(b)) are calculated from the fluid velocity field, derived by coarse-graining the particle data onto a fluid grid with cells. Here, we choose $40\times 40\times 40$ grid to ensure adequate particle averaging within each grid cell, resulting in approximately 32 particles per cell. 
\subsection{Coarse-graining particle data into fluid data}
To extract continuum-level information from particle-based LAMMPS simulations, a spatial coarse-graining procedure is employed to compute fields such as velocity and potential energy. The simulation box is divided into uniform grid bins along each spatial dimension. For each bin, local averages are calculated using particle data falling within the bin volume. The bin size of $40\times40\times40$ is chosen (which result in approximately $31$ particles per bin) to balance resolution and statistical convergence . The velocity field $\textbf{v(r)}$ at the center of each bin is computed as the mass-weighted average velocity of particles inside the bin. $\mathbf{v}(\mathbf{r}) = \frac{1}{M_{\text{bin}}} \sum_{i \in \text{bin}} m_i \mathbf{v}_i$, here $\mathbf{v}_i$ and $m_i$ is the velocity and mass of particle $i$ respectively, and $M_{\text{bin}} = \sum_{i \in \text{bin}} m_i$ is the total mass in the bin. Since all the particles have equal mass, the expression simplifies to the arithmetic mean of velocities in the bin.

\begin{figure}[!ht]
    \centering
    \includegraphics[width=0.7\textwidth]{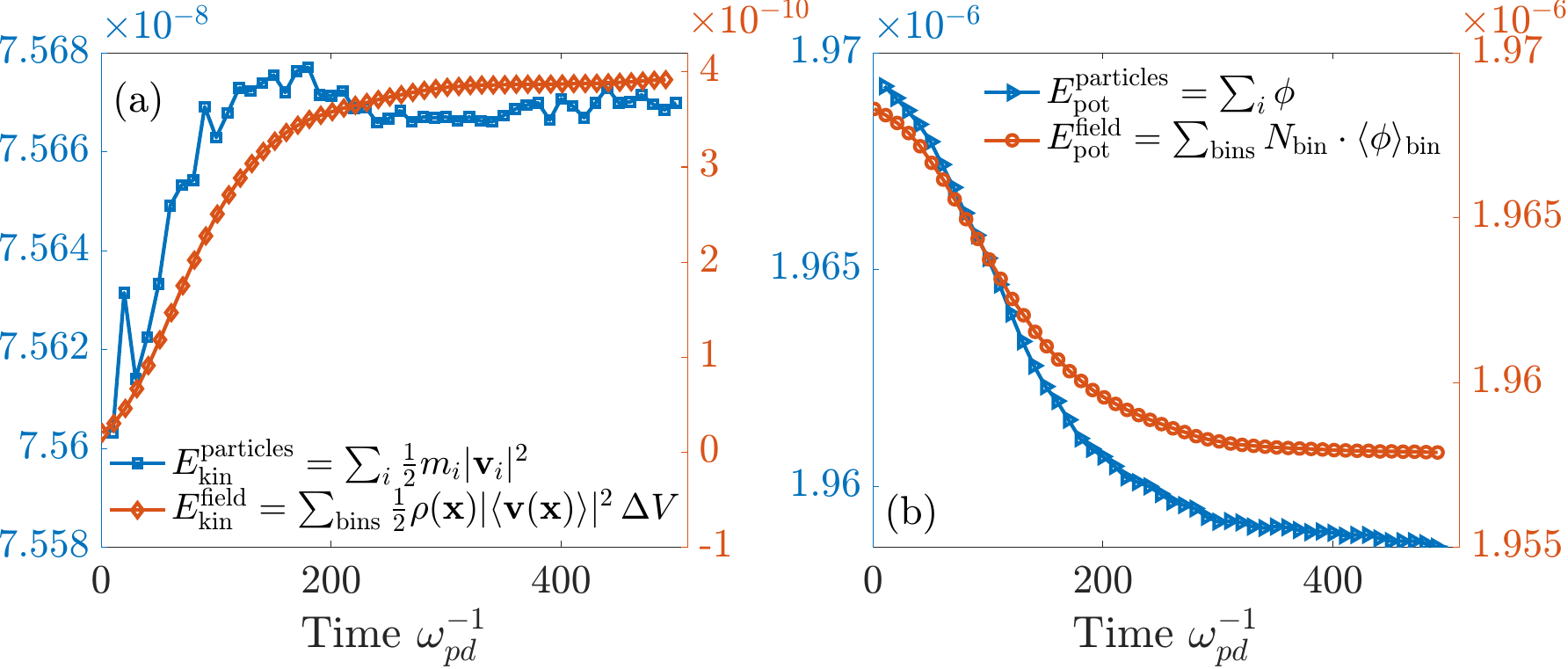}
 \caption{Time evolution of kinetic (a) and potential (b) energy computed from both particle data and coarse-grained fields. The close agreement between the two methods confirms the validity of the coarse-graining approach for capturing macroscopic energy dynamics.} 
   \label{fig:ke_pe_part_fluid}
\end{figure}

The coarse-grained potential energy field $\Phi(r)$ is computed by averaging the per-particle potential energies over all particles within each bin.$\Phi(\mathbf{r}) = \frac{1}{N_{\text{bin}}} \sum_{i \in \text{bin}} \phi_i$, here $\phi_i$ is the per-particle potential energy of particle $i$. The evolution of kinetic and potential energies at the particle and fluid level are evaluated as shown in figure~\ref{fig:ke_pe_part_fluid}. The near constancy of both energies over time indicates that the system has reached a statistically steady state, where energy input, transfer, and dissipation are balanced.
\subsection{Computation of Reynolds number and Weissenberg number}
 In our simulation, we calculated the stress tensor components ($\sigma_{xx}$, $\sigma_{yy}$, $\sigma_{xy}$ etc.) to evaluate the Weissenberg number $W_i$ and the Reynolds number $R_e$. In dusty plasmas, the stress tensor includes both kinetic contributions arising from the motion of dust particles and potential (virial) contributions due to the screened Coulomb (Yukawa) interactions between them. Physically, the stress tensor quantifies how momentum is transported and stored within the system. The kinetic component represents the momentum flux associated with particle velocities, while the potential component captures the elastic response resulting from interparticle forces. In strongly coupled dusty plasmas, the potential (elastic) stress typically dominates, allowing the system to exhibit viscoelastic behavior by sustaining shear and compressive stresses. The stress components are computed using the Irving–Kirkwood formulation~\cite{Irving_Kirkwood}, which provides a microscopic expression for the stress tensor based on particle positions, velocities, and interactions.
%\begin{equation}
%    \nonumber
$    \sigma_{\alpha\beta}=\frac{1}{V}\left [\sum_i m_iv_{i\alpha}v_{i\beta}+0.5\sum_{i\neq j}r_{ij\alpha}F_{ij\beta}\right]$
%\end{equation}
The time dependent Weissenberg number is then computed $W_i=\sigma_{\mathrm{pot}}/\sigma_{\mathrm{visc}}$ which gives a mean value of Weissenberg number to be 2.5. Here $\sigma_{\mathrm{pot}}$ is the potential (elastic) contribution to the pressure stresses, and $\sigma_{\mathrm{visc}}$ is the viscous stresses $\langle \epsilon_{\mathrm{visc}} \rangle = \nu \left\langle \sum_{i,j} \left( \frac{\partial u_i}{\partial x_j} \right)^2 \right\rangle$. The viscous stresses are computed from the coarse-grained fluid data and the inertial and elastic stresses are computed using the particle data. Similarly the value of the Reynolds number is approximately $12$. The value of viscosity for 3D dusty plasmas at $\Gamma=1$ and $\kappa=0.1$ is $1\times 10^{-9}$ Pa.s~\cite{saigo2002shear}

\section{Results and Discussions}
\subsection{Energy budget and spectral analysis}
\begin{figure}[!ht]
    \centering
    \includegraphics[width=0.6\textwidth]{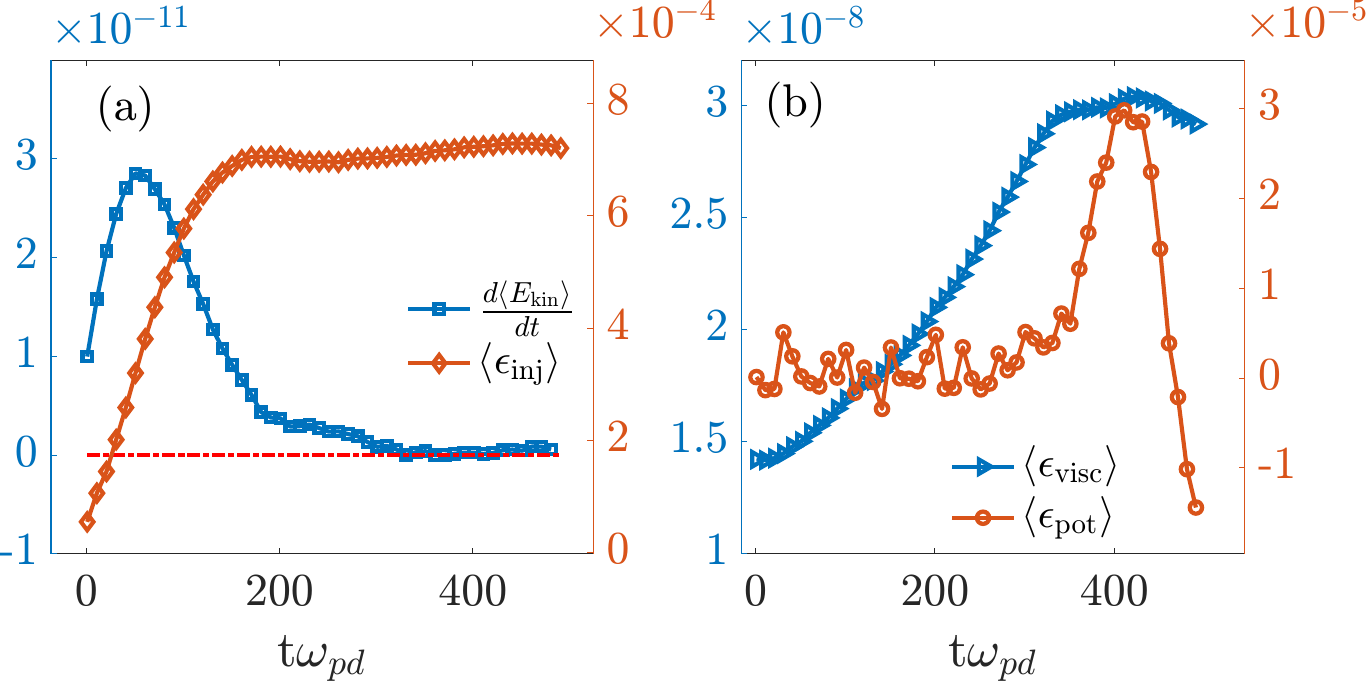}
    \caption{(a) Total kinetic energy decay rate $d\langle E_{\mathrm{kin}}\rangle/dt$, fluctuates at zero, energy injection $\langle\epsilon_{\mathrm{inj}}\rangle = \sum f(k).v(k)$ remains constant. (b) Time evolution of viscous dissipation $\langle \epsilon_{\mathrm{visc}} \rangle = \nu \left\langle \sum_{i,j} \left( \frac{\partial u_i}{\partial x_j} \right)^2 \right\rangle$  and elastic dissipation $\langle \epsilon_{\mathrm{pot}} \rangle = - \left\langle \nabla \Phi \cdot \mathbf{u} \right\rangle$.}
    \label{fig:idte}
\end{figure}

The temporal evolution of the kinetic energy budget reveals a system attaining a quasi-steady state~(fig.~\ref{fig:idte}).
At the steady state, the injection rate $\epsilon_{\mathrm{inj}}$ remains approximately constant, while the rate of change of total kinetic energy $dE_{\mathrm{kin}}/dt$ fluctuates around zero, indicating a dynamic balance between energy input and dissipation. The viscous dissipation rate is 3 orders lower in magnitude than the elastic dissipation rate in magnitude, signifying that elastic stresses play a dominant role in the energy cascade. The presence of significant elastic dissipation implies that the flow is not purely hydrodynamic, but rather exhibits strong viscoelastic character. This elastic contribution arises from the interaction of the flow with the potential energy landscape generated by the Yukawa interparticle forces, which resists deformation and interchanges the kinetic energy from the system.

A sudden transition of the elastic dissipation rate $\epsilon_{\mathrm{pot}}$ from positive to negative signifies a quick reversal in the direction of energy transfer between the flow and the elastic field.
While a positive value of the $\epsilon_{\mathrm{pot}}$ reflects a regime where the elastic field is actively releasing stored energy back into the flow a phenomenon analogous to elastic recoil. In contrast, a negative value of $\epsilon_{\mathrm{pot}}$ indicates that the flow is doing work against the potential field, resulting in storage of kinetic energy as elastic energy. This abrupt sign change typically arises from nonlinear feedback in the system: as the flow stretches or distorts the elastic field, elastic stresses build up until a critical point is reached. Beyond this point, the field can no longer sustain the deformation and snaps back, converting stored potential energy into kinetic energy.

\begin{figure}[!ht]
\centering
\includegraphics[width=0.5\textwidth]{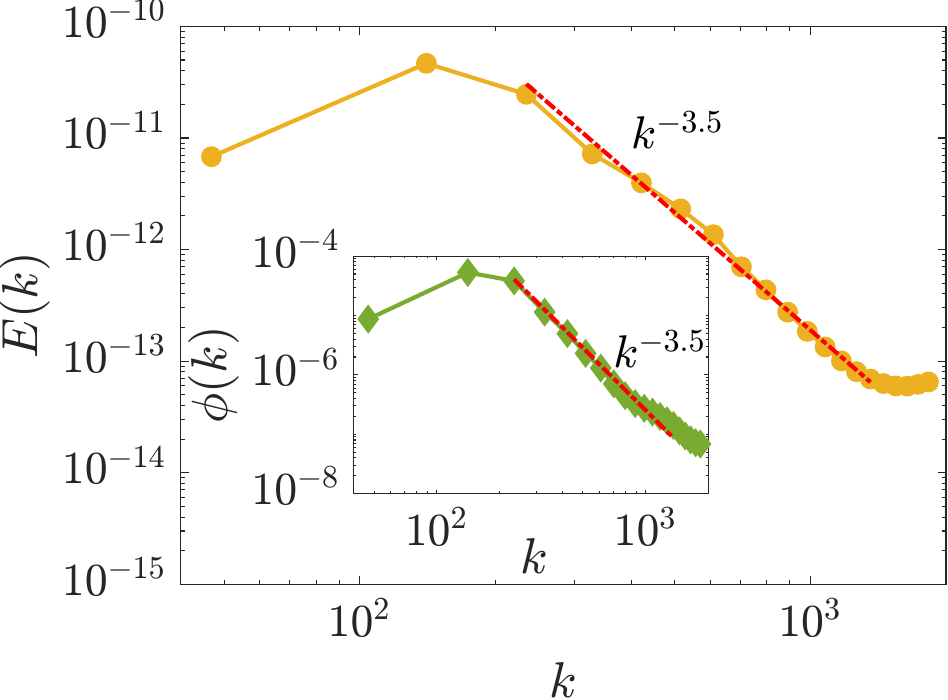}
\caption{Log-log plot of kinetic energy $E(k)$ and elastic energy $\Phi(k)$ vs wavenumber $k$. The black dashed line with slope $k^{-3.5}$ is fitted in the inertial range. The slope is steeper than the standard Kolmogorov law due to viscoelastic effects in the medium. The elastic energy spectrum (inset) also exhibits a cascade with $k^{-3.5}$ scale spectrum.} 
\label{fig:spectrum}
\end{figure}

We analyzed the kinetic energy spectrum of the velocity field in the spatial domain and observed a power-law scaling over nearly a decade of wavenumbers, as shown in fig.~\ref{fig:spectrum}. The kinetic energy spectrum is computed under steady-state conditions and exhibits an energy cascade from large to small scales, following a $k^{-3.5}$ scaling. This exponent is significantly steeper than the classical Kolmogorov theory. The steepening arises because, in our system, part of the kinetic energy is intermittently stored as elastic energy, thereby reducing the kinetic energy flux across scales. The origin of this elastic behavior lies in the interaction energy between charged particles. When this interaction energy becomes comparable to the thermal kinetic energy, it leads to structured arrangements and imparts viscoelastic, fluid-like properties to the system. A similar scaling exponent has been reported in experimental observations of viscoelastic turbulence by Groisman et al.~\cite{Groisman:Nature2000}.

Although the Reynolds number in present system is low ($R_e \sim 10$), which typically corresponds to laminar flow, the Weissenberg number is high ($W_i = 2.5$), indicating that elastic stresses dominate over viscous stresses~(see supplementary material). The elastic field not only passively stores energy but also actively interacts with and dampens the flow field, redistributing energy non-locally. To support the presence of an elastic energy cascade, we plot the elastic energy spectrum in fig.~\ref{fig:spectrum}(inset). It shows a clear transfer of elastic energy from large to small scales, following $\Phi(k) \propto k^{-3.5}$ in the inertial range. This behavior is a consequence of non-local elastic interactions, where each particle is influenced by multiple neighbors over a finite range. As a result, energy storage and release occur through collective, extended motion rather than isolated particle displacements. Consequently, elastic energy is predominantly released at intermediate and large scales, while small-scale dynamics are suppressed due to damping and thermal noise.

During the phases of positive $\epsilon_{\mathrm{pot}}$ the elastic energy enhances the kinetic activity, mainly at intermediate and larger scales. This is supported by the measured integral length scale which is computed as 
$L_{\mathrm{int}}=1/\langle v^2\rangle\int_0^{\infty} \langle v(r +dr).v(r)\rangle dr$. The integral length scale is approximately $13~\mathrm{mm}$, which is significantly smaller than the box size $67~\mathrm{mm}$, indicating that most of the kinetic energy resides in intermediate, coherent structures rather than system-wide flows. When the elastic dissipation is negative, kinetic energy is temporarily stored as elastic energy, weakening the cascade and steepening the kinetic energy spectrum. Thus, elastic modes act as a dynamic energy reservoir, intermittently modulating the cascade and driving the bursty, scale-dependent behavior typical of viscoelastic turbulence. 

\begin{figure}[H]
    \centering
    \includegraphics[width=0.8\textwidth]{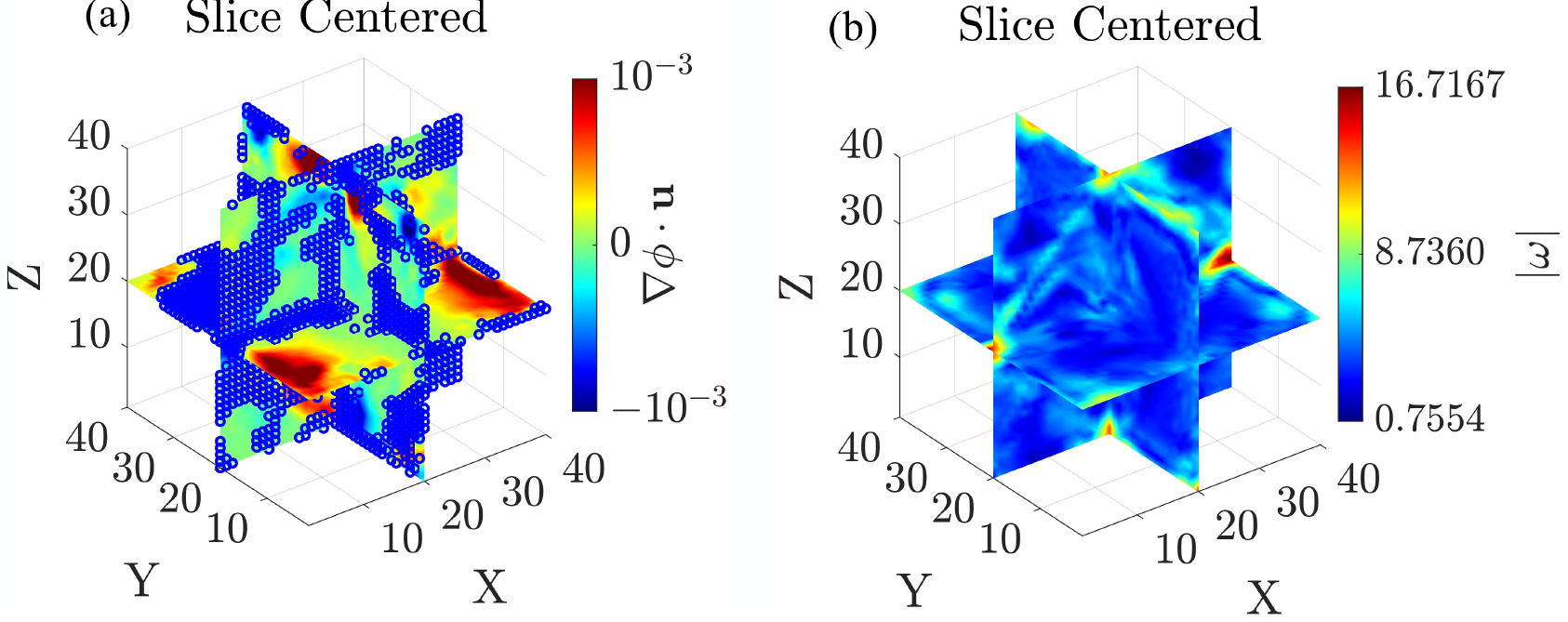}
    \caption{(a) Blue circles represent the regions where local elastic power density $\nabla \phi \cdot \mathbf{u}$ is negative (where elastic energy is released) in the YZ, XZ, and XY planes. Regions not marked by circles represent positive values of $\nabla \phi \cdot \mathbf{u}$ (where elastic energy is stored). (b) 3D slice visualization of the vortex field showing a central coherent vortex structure surrounded by localized high-vorticity regions
    }
    \label{fig:positive_negative_ed}
\end{figure}

To gain spatial insight into the elastic energy exchange dynamics, we computed and visualized the local elastic power density field $\boldsymbol{\nabla} \Phi \cdot \mathbf{u}$, which quantifies the pointwise rate of energy transfer between the elastic and velocity fields. This quantity reveals whether kinetic energy is being locally stored as elastic energy (negative values) or whether elastic energy is being released into the flow (positive values). The spatial slices (fig.~\ref{fig:positive_negative_ed}) of this field displays a complex pattern of red and blue patches, corresponding to regions of energy release and storage, respectively. 
\subsection{Intermittency in flow dynamics}
The localized zones of positive and negative elastic power are indicative of intermittent stress dynamics, suggesting that elastic energy exchange is not uniform but occurs in bursts. The vorticity plot (fig.~\ref{fig:positive_negative_ed} b) reveals the underlying rotational structure of the flow, with sharper gradients and localized high vorticity regions suggesting vortex stretching and strong shear. In SCP these vortical regions not only govern momentum transport but also stretch the elastic fields leading to energy storage. This stored energy can later be released through elastic recoil, feeding back into the flow and sustaining complex turbulent structures. We capture this intermittent dynamics using the higher order moments of velocity increments in the flow, often referred to as Lagrangian structure functions and their probability distribution functions as discussed below.
\begin{figure}[!ht]
    \centering
    \includegraphics[width=0.5\textwidth]{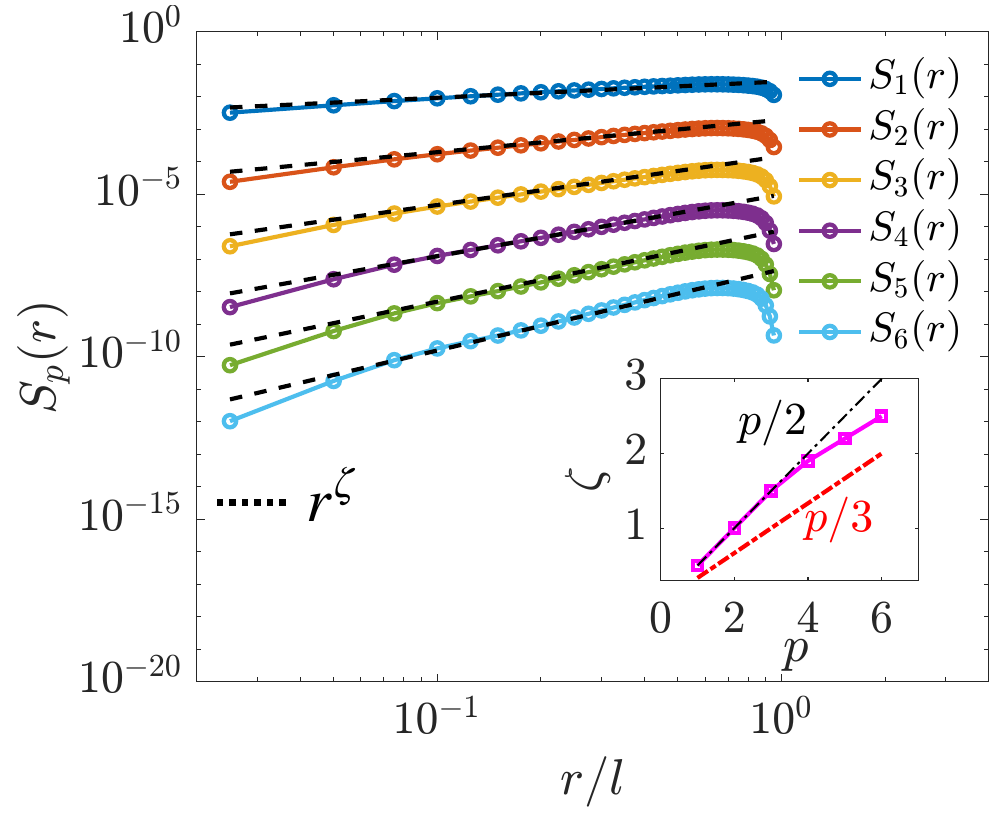}
    \caption{Structure functions $S_p(r)$ of velocity increments $\delta u_r$ up to sixth order. The exponents $\zeta$ vs $p$ calculated from the scaling relation of $S_p(r)$ shows deviation from the $p/2$ (black dashed) line reflects the intermittency. The Kolmogorv $p/3$ (red dashed) line is plotted for comparison purpose.
    }
    \label{fig:structure_function}
\end{figure}
In homogeneous isotropic turbulence (HIT), the velocity increments in the inertial range follows the scaling relation $\langle[v(x+r)-v(r)]\rangle^p=S_p(r)\propto r^{\zeta_p}$, which is universal in nature~\cite{kolmogorov1991local,Pandit:Pramana2009}. Using dimensional analysis Kolmogorov predicted that $\zeta_p \sim p/3$. However, it has been established later that the scaling exponent $\zeta_p$ is a nonlinear convex function of $p$, this phenomenon is known as intermittency~\cite{Frisch:JFM1978,Sreenivasan:JFM1985}. The exact expression of the scaling exponent $\zeta_p$ of the Navier-Stokes equations is still a central goal of turbulence research.

\begin{figure}[!ht]
    \centering
    \includegraphics[width=0.5\textwidth]{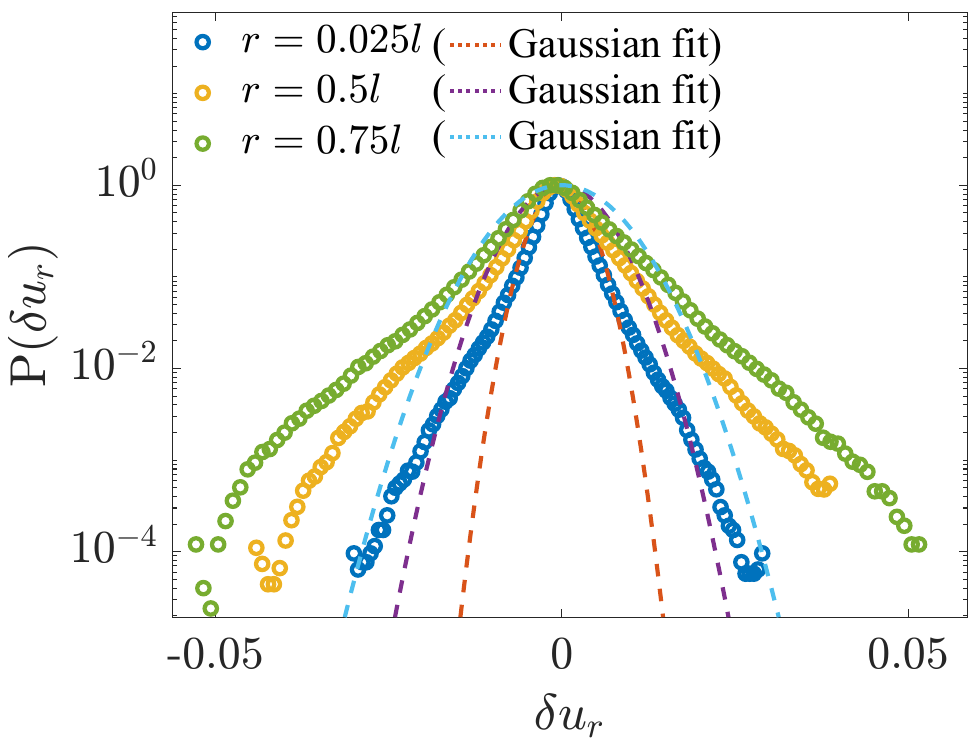}
    \caption{The probability distribution function of velocity increments $\delta u_r$ at three distant values of $r$. The distributions are non-Gaussian at the tails at all increments, reflecting intermittency}
    \label{fig:pdf_increments}
\end{figure}

It is known that the velocity field in elastic turbulence is smoother than in HIT primarily due to the dominance of elastic and viscous dissipation over inertial effects.
Here, we show that the scaling exponent $\zeta_p$ for viscoelastic dusty plasma turbulence follows $p/2$ scaling rather than the $p/3$ (Fig.~\ref{fig:structure_function} inset). Moreover, the scaling exponents also deviate from the $p/2$ line at higher orders, reflecting intermittency. To further support our results, we computed the probability distribution functions (PDFs) of longitudinal velocity increments $\delta u_r$, at various spatial separations $r$. As shown in (Fig~\ref{fig:pdf_increments}) the PDFs exhibit increasingly heavy tails as $r$ decreases, indicating strong deviations from Gaussian statistics. This non-Gaussian behavior most pronounced at small scales is a represents intermittency, where rare, intense fluctuations dominate the dynamics.

Thus, we demonstrate viscoelastic turbulence with intermittency in SCP that arises at low Reynolds numbers, driven not by inertia but by a balance between elastic stress and viscous dissipation. Through large-scale 3D molecular dynamics simulations, we observed steep kinetic energy spectra, anomalous structure function scaling, and strong intermittency, in SCP system. Although elastic turbulence in polymer solutions often leads to smooth velocity fields due to strong viscous and elastic damping at small scales, our observations in dusty plasmas, despite their viscoelastic nature reveal significant intermittency in the velocity field. This suggests that while the plasma medium behaves like a viscoelastic fluid, the underlying long-range interactions , collective dynamics, introduce complex spatiotemporal fluctuations that deviate from the smooth dynamics seen in classical elastic turbulence. Future work could explore the dependence of viscoelastic turbulence on ($\Gamma$), ($\kappa$), and system dimensionality, as well as investigate whether similar behavior can be observed in laboratory dusty plasma experiments under controlled forcing conditions.

\section{Conclusion}
Our large-scale three-dimensional molecular dynamics simulations of driven dissipative strongly coupled plasmas reveal the emergence of intermittent viscoelastic turbulence, a phenomenon driven by the intrinsic interplay between elastic and viscous dissipation at the particle scale. Unlike traditional hydrodynamic turbulence, which relies on high Reynolds numbers and inertial-viscous cascades, the flow dynamics in SCPs at low Re $\sim$ 12 and Wi $\sim$ 2.5 exhibit distinct viscoelastic characteristics. The observed power-law scalings in both kinetic and elastic energy spectra, $E(k) \propto k^{-3.5}$ and $\Phi(k) \propto k^{-3.5}$, show the viscoelastic turbulent features in SCP.

In real space, the velocity structure functions display anomalous scaling exponents, deviating from Kolmogorov-like predictions, while the probability distribution functions of velocity increments exhibit heavy-tailed, non-Gaussian behavior reflection of intermittency. 

These results establish SCPs as a new platform for probing viscoelastic turbulence at the microscopic level and provide insights relevant to astrophysical plasmas, soft matter, and nonequilibrium statistical mechanics. Future investigations could extend this framework to higher coupling strengths ($\Gamma \gg 1$)  or integrate hybrid MD-hydrodynamic approaches to scale up to experimentally accessible regimes. By unveiling the microscopic origins of viscoelastic intermittency, our results pave the way for predictive theories of energy cascades in correlated systems, enriching our understanding of turbulence beyond the Navier-Stokes paradigm.
%%%%%%%%%%%%%%%%%%%%%%%%%%%%%%%%%%%%%%%%%%%%
\section{Acknowledgements}
The authors thank Prof. Mahendra Verma and Prof. Abhijit Sen for the important discussions and the  critical reading of the manuscript. We also acknowledge the use of AGASTYA HPC for present studies and the support for this work through SERB Grant No. CRG/2020/003653.
% and MKV, the support of SERB Grant Nos. SERB/PHY/20215225 and SERB/PHY/2021473.
%%%%%%%%%%%%%%%%%%%%%%%%%%%%%%%%%%%%%%%%%%%%
% \bibliography{Euler_turb.bib,journal.bib,book.bib,preprint.bib}

%\bibliography{Euler_turb.bib}
%\section{Appendix}


%%%%%%%%%%%%%%%%%%%%%%%%%%%%%%%%%%%%%%%%%%%%
\end{document}